\documentclass{llncs}
\usepackage{amssymb}

\usepackage[centredisplay,PostScript=dvips]{diagrams}
\usepackage{mathrsfs}

\title{Parametrizing Program Analysis by Lifting to Cardinal Power Domains}
\author{Lunjin Lu}
\institute{Oakland University, MI 48309, USA}

\newcommand {\dom} [1]{{\mathit{dom}(#1)}}
\newcommand {\rename} [3]{{\mathcal{R}_{#1\to #2}(#3)}}

\def\fail{{\it fail}} 
\def\mgu{{\it mgu}}
\def\func{{\Sigma} } 
\def\allvars{{\sf Vars} }

\def\inputinfo{\mathscr{I}}
\def\outputinfo{\mathscr{O}}

\def\tlist{\mathit{list}}
\def\tnat{\mathit{nat}}
\def\tint{\mathit{int}}

\def\pars{\mathbb{P}}

\def\pos{\mathcal{P}os}
\def\con{\mathcal{G}}
\def\bool{\mathit{Bool}}
\def\mono{\stackrel{m}{\longrightarrow}}

\newcommand{\lift}[1]{\star{#1}}
\newcommand{\lifttwo}[2]{\star_{#1}{#2}}
\newcommand{\lfp}[2]{{\mathrm{lfp}_{{#1}}~#2}}
\newcommand\gc[4]{{\langle #1,#2,#3,#4\rangle}}

\newcommand{\comment}[1]{}

\def\fail{{\diamond}}

\newcommand{\vars}[1]{{{\bf V}\!(#1)}}

\def\allvars{{\cal V} }

\newcommand {\definedas}{=}

\def\term{{\sf Term}}

\def\tnat{{\it nat}}
\def\tint{{\it int}}

\def\mgu{{\it mgu}}

\def\func{\Sigma}

\def\dsub{{\it Subst}}

\begin{document}
\maketitle

\begin{abstract}
A parametric analysis is an analysis whose input and output are
parametrized with a number of parameters which can be instantiated to
abstract properties after analysis is completed.  This paper proposes
to use Cousot and Cousot's Cardinal power domain to capture functional
dependencies of analysis output on its input and obtain a parametric
analysis by parametrizing a non-parametric base analysis.  We
illustrate the method by parametrizing a $\pos$ based groundness
analysis of logic programs to a parametric groundness analysis. In
addition, a prototype implementation shows that generality of the
parametric groundness analysis comes with a negligible extra cost.
\end{abstract}

\section{Introduction} \label{sec:intro}

A program analysis is to infer information from programs. Let $P$ be a
program, $\inputinfo$ express input information before analysis, and
$\outputinfo$ express output information inferred from $P$ and
$\inputinfo$. We write $\langle \inputinfo,P, \outputinfo \rangle$ to
denote the analysis that infers $\outputinfo$ from $P$ and
$\inputinfo$.  A typical program analysis is non-parametric in the
sense that the program need be analyzed separately for different input
information. Note that program variables are not parameters for input
information, though input information can be thought of as predicates
over program variables. Take the generic sorting program $sort(x,y)$
for instance, letting $\tnat$ denote the set of natural numbers,
$\tint$ the set of integers, and $\tlist(\beta)$ the set of lists of
elements from $\beta$, program analyses $\langle
x\in\tlist(\tnat),sort(x,y), y\in\tlist(\tnat)\rangle$ and $\langle
x\in\tlist(\tint), sort(x,y), y\in\tlist(\tint)\rangle$ are
accomplished separately even if they are two instances of a parametric
analysis $\langle x\in\tlist(\beta),sort(x,y),
y\in\tlist(\beta)\rangle$ where both input information and output
information are parametrized.  By assigning different values to
$\beta$ which serves as a place holder for information to be filled in
after analysis, $\langle x\in\tlist(\beta), sort(x,y),
y\in\tlist(\beta)\rangle$ can be instantiated into many different
non-parametric analyses such as $\langle x\in\tlist(\tnat),sort(x,y),
y\in\tlist(\tnat)\rangle$ and $\langle x\in\tlist(\tint),sort(x,y),
y\in\tlist(\tint)\rangle$.  Parametric program analyses infer more
general results, which brings some benefits. Firstly, a sub-program or
a library program need not be analyzed separately for its different
uses, i.e., the result of a parametric analysis is re-usable. This has
positive bearing on efficiency of analysis because output information
for different uses of the same sub-program can be obtained by
instantiation rather than by re-analysis. Secondly, parametric
analyses are amenable to program modifications since changes to the
program does not necessitate re-analyses of the sub-program so long as
the sub-program itself is not changed.

This paper addresses the issue of lifting a non-parametric analysis to
a parametric analysis such that each instantiation of the result of
running the parametric analysis is same as the result of running the
non-parametric analysis with instantiated input information. If
$\langle \inputinfo(\vec{\beta}), P, \outputinfo(\vec{\beta})\rangle$
is the result of the parametric analysis then $\langle
\inputinfo(\kappa), P, \outputinfo(\kappa)\rangle$ is the result of
the non-parametric analysis for any possible value $\kappa$ for
$\vec{\beta}$.  Observe that both input $\inputinfo(\vec{\beta})$ and
output $\outputinfo(\vec{\beta})$ of the parametric analysis are
functions from the domain of values for parameters to the domain of
input properties non-parametric analysis.

The contributions of the paper are as follows.  Firstly, a systematic
approach is presented for deriving a parametric analysis from a given
non-parametric base analysis. This involves lifting the semantic
domain for the base analysis to its Cardinal power with respect to the
domain of parameter values and lifting the semantic function
accordingly. Secondly, this approach is applied to a goal-dependent
groundness analysis for logic programs using parameters to express
groundness of variables in the top-level goal. The result is a
parametric goal dependent groundness analysis. Thirdly, we present an
encoding for the abstract properties and abstract operations for the
parametric groundness analysis using positive propositional
formulas. 

The next section provides background knowledge on abstract
interpretation and logic program analysis. Section~\ref{sec:lift}
describes the approach to parametrizing program analyses and
section~\ref{sec:para} obtains the parametric goal-dependent
groundness analysis for logic programs by applying the
approach. Section~\ref{sec:encode} presents the encoding and
section~\ref{sec:imp} some experimental results with a prototype
implementation of the parametric groundness analysis. We discuss
related work in section~\ref{sec:related} and then conclude in
section~\ref{sec:conc}.

\section{Preliminaries} \label{sec:pre}

\subsection{Abstract Interpretation}

A semantics of a program is given by an interpretation $\langle
(D,\sqsubseteq_D),f\rangle$ where $(D,\sqsubseteq_D)$ is a complete
lattice and $f$ is a monotone function on $(D,\sqsubseteq_D)$. The
semantics is defined as the least fixed point $\mathrm{lfp}~{f}$ of
${f}$. The concrete semantics of the program is given by the concrete
interpretation $\langle (D,\sqsubseteq_D),{f}\rangle$ while an
abstract semantics is given by an abstract interpretation $\langle
(D^\sharp,\sqsubseteq_D^\sharp),{f^\sharp}\rangle$.  The
correspondence between the concrete and the abstract domains is
formalized by a Galois connection $(\alpha,\gamma)$ between
$(D,\sqsubseteq_D)$ and $(D^\sharp,\sqsubseteq_D^\sharp)$.  A Galois
connection between $D^\sharp$ and $D$ is a pair of monotone functions
$\alpha:{D}\mapsto{D^\sharp}$ and $\gamma:{D^\sharp}\mapsto{D}$
satisfying $\forall c\in{D}.  (c\sqsubseteq_D \gamma\circ\alpha(c))$
and $\forall {a}\in{D^\sharp}.
(\alpha\circ\gamma({a})\sqsubseteq_D^\sharp{a})$. The function
$\alpha$ is called an abstraction function and the function $\gamma$ a
concretization function.  A sufficient condition for
$\mathrm{lfp}{{f^\sharp}}$ to be a safe abstraction of
$\mathrm{lfp}~{{f}}$ is
$\forall{a}\in{D^\sharp}.(\alpha\circ{f}\circ\gamma({a})~\sqsubseteq_D^\sharp~{f^\sharp}({a}))$
or equivalently
$\forall{a}\in{D^\sharp}.({f}\circ\gamma({a})~\sqsubseteq_D~\gamma\circ{f^\sharp}({a}))$,
according to propositions 24 and 25 in~\cite{Cousot:JLP92}.  \comment
{ The abstraction and concretization functions in a Galois connection
  uniquely determine each other; and a complete meet-morphism
  $\gamma:{D^\sharp}\mapsto{D}$ induces a Galois connection
  $(\alpha,\gamma)$ with $\alpha(c)=\sqcap_{D^\sharp}
  \{{a}~|~c\sqsubseteq_D\gamma({a})\}$. A function
  $\gamma:{D^\sharp}\mapsto{D}$ is a complete meet-morphism iff
  $\gamma(\sqcap_{D^\sharp} X)=\sqcap_{D}\{\gamma(x)\in{X}\}$ for any
  $X\subseteq{D^\sharp}$. Thus, an analysis can be formalized as a
  tuple \( (\langle (D,\sqsubseteq_D),{f}\rangle,\gamma,\langle
  (D^\sharp,\sqsubseteq_D^\sharp),{f^\sharp}\rangle)\) such that
  $\langle (D,\sqsubseteq_D),{f}\rangle$ and $\langle
  (D^\sharp,\sqsubseteq_D^\sharp),{f^\sharp}\rangle$ are
  interpretations, $\gamma$ is a complete meet-morphism from
  $({D},\sqsubseteq_D)$ to $({D^\sharp},\sqsubseteq_D^\sharp)$, and
  $\forall{a}\in{D^\sharp}.({f}\circ\gamma({a})~\sqsubseteq_D~\gamma\circ{f^\sharp}({a}))$.
} In a compositional design of analysis, the concrete semantics is
defined in terms of a group of semantic functions $f_{i}:D_{i}\mapsto
E_{i}$ and the abstract semantics is defined in terms of another group
of semantic function $f_{i}^\sharp:D_{i}^\sharp\mapsto E_{i}^\sharp$
such that each abstract semantic function $f_{i}^\sharp$ simulates its
corresponding concrete semantic function $f_{i}$. To prove the
correctness of the abstract semantics with respect to the concrete
semantics is reduced to proving the correctness of each abstract
semantic function $f_{i}^\sharp$ with respect to its corresponding
concrete semantic function $f_{i}$.  Let $\gamma_{D_i^\sharp}:
D_i^\sharp\mapsto D_i$ and
$\gamma_{E_i^\sharp}:E_{i}^\sharp\mapsto{E}_{i}$ be concretization
functions. Then $f_{i}^\sharp:D_{i}^\sharp\mapsto E_{i}^\sharp$ is
correct with respect to $f_{i}:D_{i}\mapsto E_{i}$ iff
$f_{i}(\gamma_{D_i^\sharp}(x^\sharp))\sqsubseteq_{E_{i}}
\gamma_{E_i^\sharp}(f_{i}^\sharp(x^\sharp))$ for each $x^\sharp\in
D_{i}^\sharp$.

\subsection{Logic Program Analysis}
 Assume a set of {function symbols} $\func$ and an infinite set of
 variables $\allvars$.  Let $V\subseteq\allvars$. Then
 $\term(\func,V)$ denotes the set of all {terms} that can be
 constructed from $\func$ and $V$.  Let $\vars{o}$ be the set of
 variables in a syntactic object $o$.  A bold lower case letter
 denotes a sequence of different variables.  When there is no
 ambiguity, $\vars{\vec{x}}$ will be written as $\vec{x}$.
The set of subsets of a set $S$ is denoted by $\wp(S)$ and the set of
finite subsets of $S$ by $\wp_f(S)$.

\paragraph{Substitutions}
A substitution $\theta$ is a mapping from $\allvars$ to
$\term(\func,\allvars)$ such that its domain $dom(\theta)=\{x\mid
x\neq\theta(x)\}$ is finite.  A substitution $\theta$ is idempotent
iff $\theta(\theta(x))=\theta(x)$ for all $x\in\allvars$. The set of
all idempotent substitutions is denoted $\dsub$. A renaming
substitution is a bijection from $\allvars$ to $\allvars$. Let
$\sim_U$ be the equivalence relation defined $\theta\sim_U\sigma$ iff
there is a renaming substitution $\rho$ such that
$\theta(x)=\rho(\sigma(x))$ for all $x\in U$~\cite{AmatoS09}.
\comment{Note that $\sim_U$ is weaker than the usual equivalence
  relation in which $\theta$ and $\rho\circ\sigma$ must agree on all
  variables in $\allvars$.}  Let $[\theta]_{U}$ denote the equivalence
class of $\theta$ with respect to $\sim_U$ and $\dsub_{U}$ the
quotient set of $\dsub$ with respect to $\sim_U$. A substitution
$\theta'$ is a canonical representative of an equivalence class
$[\theta]_U$ iff $\theta'\in [\theta]_U$ and $\dom{\theta'}=U$ and
$U\cap rng(\theta')=\emptyset$ where $rng(\theta')=\cup_{x\in
  dom(\theta')}\vars{\theta'(x)}$. In any $\theta'$ in $[\theta]_U$,
bindings for variables outside U is irrelevant since $\{x\mapsto
\theta'(x)\mid x\in U\} \sim_{U} \theta'$.  \comment{ We will use
  $\eta$ and $\zeta$ to denote equivalence classes of substitutions
  with respect to $\sim_U$ for some $U$ that is either irrelevant or
  clear from the context.}  Let $\fail\not\in\dsub_{U}$ for any
$U\subseteq\allvars$.

\paragraph{Operations on substitutions}
An equational constraint is a finite set (conjunction) of equations of
the form $t_1=t_2$ with $t_i$ for $i=1,2$ being terms.  Define
$\mgu(E)$ as the $\sim_{\vars{E}}$ equivalence class of most general
unifiers for $E$ if $E$ is unifiable. Otherwise, $\mgu(E)=\fail$.  One
operation performed during program execution is to conjoin constraints
represented by substitutions. The unification operation
$\odot:\dsub_{U}\times\dsub_{V}\mapsto \dsub_{U\cup V}\cup\{\fail\}$
is defined by \({[\theta_1]}_{U} \odot {[\theta_2]}_{V}
=\mgu(eq(\theta_1')\cup eq(\theta_2')) \) where
$eq(\theta)=\{x=\theta(x)\mid x\in dom(\theta)\}$ and $\theta_1'$ and
$\theta_2'$ are respectively canonical representatives of
${[\theta_1]}_{U}$ and ${[\theta_2]}_{V}$ such that
$(U\cup\vars{\theta_1'})\cap(V\cup\vars{\theta_2'})\subseteq U\cap
V$. Another operation is projection
$\pi_X:\dsub_{U}\mapsto\dsub_{U\setminus X}$ for $X\in\wp_f(\allvars)$
defined as $\pi_X([\theta]_U) = [\theta]_{U\setminus X}$. The operator
$\pi_X$ hides variables in $X$. A third operation is renaming defined
as follows. If $\vec{x}\cap\vec{y}=\emptyset$ then
$\rename{\vec{x}}{\vec{y}}{\eta} = \pi_{\vec{x}}(
\mgu(\{\vec{x}=\vec{y}\}) \odot \eta )$. Otherwise, \(
\rename{\vec{x}}{\vec{y}}{\eta} =
\rename{\vec{z}}{\vec{y}}{\rename{\vec{x}}{\vec{z}}{\eta}}\) where
$\vec{z}\cap (\vec{x}\cup\vec{y}\cup\vars{\eta})=\emptyset$. 
Note that $\mgu(\{\vec{x}=\vec{y}\})\neq\fail$ since 
$\vec{x}$ and $\vec{y}$ are sequences of different variables.
$\rename{\vec{x}}{\vec{y}}{\cdot}$ transforms an equational constraint
on $\vec{x}$ to one on $\vec{y}$.

\paragraph{Concrete and Abstract Interpretations}
The concrete semantics for a logic program analysis is usually defined
in terms of several operations on the collecting domains
$\Pi_{U\in\mathcal{U}} \langle\wp(\dsub_U),\subseteq\rangle$ where
each $U\in\mathcal{U}$ is the set of program variables of interest at
a specific program point and $\mathcal{U}$ represents the set of
program points of interest. The concrete interpretation is 
\[  I = \langle \Pi_{U\in\mathcal{U}} \langle\wp(\dsub_U),\subseteq,\cup\rangle, 
\odot^\star,  \mathcal{R}^\star_{\vec{x}\mapsto\vec{y}}, \pi^\star_{X} 
 \rangle
\]
where $\cup$ is the set union and $\odot^*,
\mathcal{R}^*_{\vec{x}\mapsto\vec{y}}$ and $\pi_{X}^*$ are set
extensions of $\odot, \mathcal{R}_{\vec{x}\mapsto\vec{y}}$ and
$\pi_{X}$ respectively.  An analysis is obtained by designing an 
abstract interpretation 
\[  I^\sharp = \langle 
 \Pi_{U\in\mathcal{U}} \langle ASub^\sharp_U,\sqsubseteq^\sharp_U,\sqcup^\sharp_U\rangle, 
\odot^\sharp,  \mathcal{R}_{\vec{x}\mapsto\vec{y}}^\sharp, \pi_{X}^\sharp  \rangle
\] such that  $\langle ASub^\sharp_U,\sqsubseteq^\sharp_U,\sqcup^\sharp_U\rangle$ 
is related to $ \langle\wp(\dsub_U),\subseteq,\cup\rangle$ with a
Galois connection
$\gc{\wp(\dsub_U)}{\alpha_U}{ASub^\sharp_U}{\gamma_U}$ and
$\odot^\sharp, \mathcal{R}_{\vec{x}\mapsto\vec{y}}^\sharp$ and
$\pi_{X}^\sharp$ approximate correctly $\odot^*,
\mathcal{R}^*_{\vec{x}\mapsto\vec{y}}$ and $\pi_{X}^*$ respectively.

\section{Parametrizing Program Analyses} \label{sec:lift}
An analysis $\langle\inputinfo, P, \outputinfo\rangle$ computes
$\outputinfo$ as the limit of an ultimately stationary sequence of
iterates: \( \inputinfo, f( \inputinfo), \cdots,
f^{(\lambda)}(\inputinfo), \cdots \) where $f$ is a monotone semantic
function. The iteration of $f$ is defined
\begin{eqnarray*}
 f^{(0)} & = & id \\
f^{(\lambda +1)} &=& f\circ f^{(\lambda)} \\
f^{(\lambda)} &=& \sqcup_{\beta<\lambda} f^{(\beta)} ~\mbox{when $\lambda$ is a limit ordinal}
\end{eqnarray*}
The limit is denoted $\lfp{\inputinfo}{f}$. As an example, consider
the forward collecting semantics which characterizes the set of the
program states $\outputinfo$ that can be reached from a set of initial
states $\inputinfo$. According to proposition 33 in
~\cite{Cousot:JLP92}, $\outputinfo = \lfp{\emptyset}{F[\![P]\!]}$ where
$F[\![P]\!](X)=\inputinfo \cup post[\stackrel{P}{\longrightarrow}]X$
and $\stackrel{P}{\longrightarrow}$ is the transition relation between
program states and \(post[t]S = \{s'\mid \exists s\in S. \langle
s,s'\rangle \in t\}\). It is easy to verify that $\outputinfo =
\lfp{\inputinfo}{\lambda X. (X\cup
  post[\stackrel{P}{\longrightarrow}]X)}$.

\subsection{Lifting Semantic Domains to Cardinal Power Domains} 
The cardinal power $L_1\mono L_2$ with base $L_2$ and exponent $L_1$
consists of all monotone functions from $L_1$ to $L_2$. \comment{  In the
classical abstract interpretation theory, Galois connections are
lifted to function space as follows. Let
$\gc{L}{\alpha}{L^\sharp}{\gamma}$ be a Galois connection. Then
\[\gc{L\mono L}{\lambda \phi.\alpha\circ\phi\circ\gamma}{L^\sharp\mono
  L^\sharp}{\lambda\psi.\gamma\circ\psi\circ\alpha}
\] is a Galois connection
(Proposition 21 \cite{Cousot:JLP92}).} We parametrize a base
analysis by lifting both the concrete and the abstract domains to cardinal powers. 
\begin{proposition}\cite{CC92jlc} \label{prop} 
Let $\gc{L_1}{\alpha_1}{L_1^\sharp}{\gamma_1}$ and $\gc{L_2}{\alpha_2}{L_2^\sharp}{\gamma_2}$ be Galois connections. 
Then
$\gc{L_1\mono L_2}{\alpha}{L_1^\sharp\mono\L_2^\sharp}{\gamma}$ is a Galois connection
where $\alpha=\lambda\phi.\alpha_2\circ\phi\circ\gamma_1$ and $\gamma=\lambda\psi.\gamma_2\circ\psi\circ\alpha_1$. 
{
\begin{proof} 
For any $\phi\in {L_1\mono L_2}$, $\phi\sqsubseteq
\gamma_2\circ\alpha_2\circ\phi\circ\gamma_1\circ\alpha_1 \sqsubseteq
\gamma\circ\alpha(\phi)$ since $\gamma_i\circ\alpha_i$ are extensive
for $i=1,2$. For any $\psi\in {L_1^\sharp\mono\L_2^\sharp}$,
$\alpha\circ\gamma(\psi) =
\alpha_2\circ\gamma_2\circ\psi\circ\alpha_1\circ\gamma_1 \sqsubseteq
\psi$ since $\alpha_i\circ\gamma_i$ are reductive for $i=1,2$. Since
$\gamma\circ\alpha$ is extensive and $\alpha\circ\gamma$ is reductive,
$\gc{L_1\mono L_2}{\alpha}{L_1^\sharp\mono\L_2^\sharp}{\gamma}$ is a
Galois connection. \qed
\end{proof}
}
\end{proposition}

\subsection{Lifting Semantic Functions}
The domain of an interpretation is often formed from a number of
primitive domains and the semantic function from a number of primitive
functions between primitive domains. We now define a family of
operators $\star_L$ that lift a monotone function $f:D\mono E$ to a
monotone function $\lifttwo{L}{f}: (L\mono D)\mono(L\mono E)$.
\begin{definition} Let $f:D\mono E$. Define $\lifttwo{L}{f}:  (L\mono D)\mono(L\mono E)$ as 
\[ \lifttwo{L}{f} = \lambda \phi. f\circ\phi
\] \label{df:lift}
\end{definition} 

The following theorem shows that lifting of the semantic function of
an interpretation can be accomplished by lifting individual primitive
semantic functions.
\begin{theorem} 
For any $L$, 
\begin{enumerate}
\item \(\lifttwo{L}{(f_2\circ f_1)} = (\lifttwo{L}{f_2}) \circ (\lifttwo{L}{f_1})\) for any $f_1:D\mono E$ and $f_2:E\mono F$,
\item  \(\lifttwo{L}{\langle f_1,f_2\rangle} = \langle \lifttwo{L}{f_1}, \lifttwo{L}{f_2}\rangle\) for any $f_1:D\mono E$ and $f_2:D\mono F$,
\item  $\lifttwo{L}{proj_i}(\langle \phi_1,\phi_2\rangle) = \phi_i$ for $i=1,2$, 
$\phi_1:L\mono D_1$ and $\phi_2:L\mono D_2$ where $proj_i(\langle c_1,c_2\rangle) = c_i$ for $i=1,2$.
\end{enumerate}

{
\begin{proof} 
Consider item (1) first. For any $\phi\in L\mono D$, 
$\lifttwo{L}{(f_2\circ f_1)}(\phi) = f_2\circ f_1\circ \phi = 
f_2\circ (\lifttwo{L}{f_1})(\phi) = (\lifttwo{L}{f_2})( (\lifttwo{L}{f_1})(\phi)) 
= (\lifttwo{L}{f_2})\circ (\lifttwo{L}{f_1})(\phi)$.

Now consider item (2). For any $\phi\in L\mono D$, 
\(\lifttwo{L}{\langle f_1,f_2\rangle}(\phi) = 
\langle f_1,f_2\rangle \circ \phi = \langle f_1\circ\phi,f_2\circ\phi\rangle 
=  \langle \lambda\psi_1.f_1\circ\psi_1,\lambda\psi_2.f_2\circ\psi_2\rangle(\phi) =  \langle \lifttwo{L}{f_1}, \lifttwo{L}{f_2}\rangle(\phi)\).

Item (3) follows from definition of $\pi_i$ and $\lifttwo{L}{\cdot}$. 
 \qed
\end{proof}
}
\end{theorem}

Let $\gc{L_2}{\alpha_2}{L_2^\sharp}{\gamma_2}$ be a Galois connection
and $f:L_2\mono L_2$ and $f^\sharp: L_2^\sharp \mono L_2^\sharp$ the
concrete and abstract semantic functions.  The concrete and abstract
domains $L_2$ and $L_2^\sharp$ can be parametrized by $L_1$ and
$L_1^\sharp$ which are related to each other by a Galois connection
$\gc{L_1}{\alpha_1}{L_1^\sharp}{\gamma_1}$.  The following theorem
says that $\lifttwo{L_1^\sharp}{f^\sharp}$ approximates
$\lifttwo{L_1}{f}$ if $f^\sharp$ approximates $f$. Furthermore, if
$f^\sharp$ is the best approximation of $f$ and
$\gc{L_1}{\alpha_1}{L_1^\sharp}{\gamma_1}$ is a Galois insertion then
$\lifttwo{L_1^\sharp}{f^\sharp}$ is the best approximation of
$\lifttwo{L_1}{f}$.

\begin{theorem} 
Let $\gc{L_1}{\alpha_1}{L_1^\sharp}{\gamma_1}$ and
$\gc{L_2}{\alpha_2}{L_2^\sharp}{\gamma_2}$ be Galois connections, 
$f: L_2\mono L_2$ and $f^\sharp: L_2^\sharp\mono L_2^\sharp$. Let $\alpha$ and $\gamma$ be defined as in Proposition~\ref{prop}.  Then 
\begin{enumerate}
\item If $\alpha_2\circ f\circ \gamma_2 \sqsubseteq f^\sharp$,
  $\alpha\circ (\lifttwo{L_1}{f})\circ \gamma \sqsubseteq
  \lifttwo{L_1^\sharp}{f^\sharp}$.
\item If $\alpha_2\circ f\circ \gamma_2 = f^\sharp$ and
  $\gc{L_1}{\alpha_1}{L_1^\sharp}{\gamma_1}$ is a Galois insertion,
  $\alpha\circ (\lifttwo{L_1}{f})\circ \gamma =
  \lifttwo{L_1^\sharp}{f^\sharp}$.
\end{enumerate}
{
\begin{proof} 
Consider  (1) first. 
Let $\phi$ be an arbitrary member of $L_1^\sharp\mono L_2^\sharp$.
\[\begin{array}{rclr}
(\alpha\circ(\lifttwo{L_1}{f})\circ\gamma)(\phi) &=& \alpha((\lifttwo{L_1}{f})(\gamma(\phi)))&\mbox{by def. of $\gamma$}\\
  &=& \alpha((\lifttwo{L_1}{f})(\gamma_2\circ\phi\circ\alpha_1))&\mbox{by def. of $\lifttwo{L_1}{f}$}\\
  &=& \alpha(f\circ\gamma_2\circ\phi\circ\alpha_1) &\mbox{by def. of $\alpha$}\\
 &=& \alpha_2\circ f\circ\gamma_2\circ\phi\circ\alpha_1\circ\gamma_1&\mbox{since $\alpha_2\circ f\circ\gamma_2\sqsubseteq f^\sharp$} \\
 &\sqsubseteq&  f^\sharp\circ\phi\circ\alpha_1\circ\gamma_1&\mbox{since $\alpha_1\circ\gamma_1$ is reductive} \\
 &\sqsubseteq&  f^\sharp\circ\phi&\mbox{by def. of $\lifttwo{L_1^\sharp}{f^\sharp}$} \\
&=& (\lifttwo{L_1^\sharp}{f^\sharp})(\phi)
\end{array}
\]
Hence, $\alpha\circ(\lifttwo{L_1}{f})\circ\gamma\sqsubseteq\lifttwo{L_1^\sharp}{f^\sharp}$.

Now consider (2). When $\alpha_2\circ f\circ \gamma_2 = f^\sharp$ and
$\alpha_1\circ\gamma_1$ is the identity function, $\sqsubseteq$
becomes $=$ in the proof for (1).  \qed
\end{proof}
}
\end{theorem}

The following result states that performing the parametric analysis
with a parametrized input and then binding the parameters to abstract
properties yields the same result as the base analysis performed with
the instantiation of the input with the same binding.

\begin{theorem} 
Let $f: D\mono D$ and $\kappa: L\mono D$. Then, for any $\ell\in{L}$, 
\[ \lfp{\kappa(\ell)}{f} =(\lfp{\kappa}{(\lifttwo{L}{f})})(\ell)
\]
{
\begin{proof} 
\( (\lfp{\kappa}{(\lifttwo{L}{f})})(\ell) = (\bigsqcup_{\beta} (\lifttwo{L}{f})^{(\beta)}(\kappa))(\ell) = (\bigsqcup_{\beta} f^{(\beta)}\circ\kappa)(\ell)
= \bigsqcup_{\beta} (f^{(\beta)}\circ\kappa)(\ell) = 
\bigsqcup_{\beta} (f^{(\beta)}(\kappa(\ell)) = \lfp{\kappa(\ell)}{f} 
\).
\end{proof} 
}
\end{theorem}

\begin{remark} In fact, any fixpoint of $\lifttwo{L}{f}$ provides 
a set of fixpoints of $f$.  Let $f: D\mono D$ and $\kappa: L\mono D$
such that $\kappa = (\lifttwo{L}{f})(\kappa)$. Then, for any
$\ell\in L$, $\lfp{\kappa(\ell)}{f} = \kappa(\ell)$ since 
\( f(\kappa(\ell)) = f\circ\kappa(\ell) = ((\lifttwo{L}{f})(\kappa))(\ell) 
=\kappa(\ell)\). 
\end{remark}

\section{Parametrizing Groundness Analysis} \label{sec:para}
In logic programming, a value is a term that may contain variables. In
any program state during the execution of a logic program, logic
variables are bound to terms that may be in turn bound to other terms
later during execution.  A variable is ground in a substitution
(program state) if the substitution maps the variable to a term that
does not contain any variable.  Groundness analysis is one of the most
studied properties for logic
programs~\cite{AMSS98,CD95:prop,CortesiFW_JLP96,Dart91,ms93,Scozzari02}.
This section present a parametric groundness  analysis by
parametrizing the groundness  analysis using positive
propositional formulas with the simplest groundness domain.

\subsection{Propositonal Formulas} 
Let $U$ be a finite set of propositional variables. A propositonal
formula over $U$ is formed of propositional constants $0$ and $1$,
propositional variables from $U$ and logical connectives $\wedge$,
$\vee$, $\leftrightarrow$ and $\neg$.  Other connectives such as
$\rightarrow$ and $\leftarrow$ can be defined using these connectives.
Let $\bool=\{0,1\}$ ordered by $0\leq 1$. A truth substitution $m$ on
domain $U$ is a partial function from $U$ to $\bool$. The application
of $m$ to $f$ is denoted $m(f)$. Let $m=\{x\mapsto 1\}$ and
$f=(x\rightarrow y)$. Then $m(f) = (1\to y)$. If a truth substitution
$m$ is defined for every propositional variables in a propositional
formula $f$ then $m$ is called a truth assignment for $f$.  Given a
formula $f$ and a truth assignment $m$, $m\models f$ means that $m$
satisfies $f$ and $f_1\models f_2$ means that $m\models f_1$ implies
$m\models f_2$ for every truth assignment $m$ for $f_1$.  Two formulas
$f_1$ and $f_2$ are equivalent, denoted $f_1= f_2$ if both $f_1\models
f_2$ and $f_2\models f_1$. We shall not distinguish between elements
in an equivalence class of $=$. A propositional formula $f$ is
positive if $\mathbf{u}\models f$ for each such truth substitution
$\mathbf{u}$ that assigns $1$ to all the propositional variables in
$f$.

\subsection{Groundness  analysis} 
Marriott and Sondergaard~\cite{ms93} proposed to use positive
propositional formulas to capture groundness dependencies between
variables in a program state. Let $x,y\in V$. Then the formula $x$
describes those program states in which $x$ is bound to a ground term
while $x\rightarrow y$ describes those program states in which $y$ is
ground whenever $x$ is.  Let $\pos_V$ denotes the set of positive
propositional formulas over propositional variables in $V$. Then
$\langle \pos_V,\models\rangle$ is a complete lattice with bottom
$\wedge V$, top $1$, meet $\wedge$ and join $\vee$.  Let $ground_V$
be defined $ground_V(\theta)=\lambda x\in
V. (\vars{\theta(x)}=\emptyset)$ and
\begin{eqnarray*} 
\alpha_{{\pos}_V}(\Theta) &=& \bigvee_{\theta\in\Theta} \exists_{-V}. \bigwedge_{x\in dom(\theta)} (x\leftrightarrow \wedge \vars{\theta(x)}) \\
\gamma_{{\pos}_V}(f) &=& \{ \theta \mid (ground_V(\theta) \models f \}
\end{eqnarray*}
 Then 
$\gc{\wp(\dsub_V)}{\alpha_{{\pos}_V}}{ \pos_{V}}{\gamma_{{\pos}_V}}$
is a Galois insertion~\cite{CortesiFW_JLP96}. Thus, the least upper bound
$\vee$ on $\langle \pos_V,\models\rangle$ approximates correctly
$\cup$ on $\langle\wp(\dsub_V),\subseteq\rangle$. The other abstract
operations for groundness  analysis are given as follows.
The abstract projection operation $\pi_{X}^\sharp: \pos_V\mapsto
\pos_{V\setminus X}$ is $\pi_{X}^\sharp(f) = \exists x_1.\exists
x_2.\cdots \exists x_n. f$ when $X=\{x_1,x_2,\cdots, x_n\}$; the
abstract unification operation
$\odot^\sharp:\pos_U\times\pos_V\mapsto\pos_{U\cup V}$ is $f_U
\odot^\sharp f_V = f_U\wedge f_V$ and the abstract renaming operation
$\mathcal{R}^\sharp_{\vec{x}\mapsto\vec{y}}:\pos_V\mapsto\pos_{V\setminus\vec{x}\cup\vec{y}}$
is defined $\mathcal{R}^\sharp_{\vec{x}\mapsto\vec{y}}(f) = f'$ where
$f'$ is obtained by simultaneously replacing the elements of $\vec{x}$
with their corresponding elements in $\vec{y}$. For instance,
$\mathcal{R}^\sharp_{{x_1x_2}\mapsto{x_2x_1}}(x_1\rightarrow x_2) =
(x_2\rightarrow x_1)$. The soundness of these operations are well
established (see, e.g. ~\cite{AMSS98}).

\subsection{Abstract domain $\con_\pars$} 
Jones and Sondergaard \cite{JS87} proposed an abstract domain that
capture groundness information in a substitution in terms of the
collection of the variables that are grounded by the substitution. Let
$\pars$ be the set of variables of interest. The above abstract domain
is isomorphic to the set of conjunctive propositional formulae with
propositional variables from $\pars$
\begin{eqnarray*} 
\con_\pars & = & \{\wedge X \mid X\subseteq \pars\}
\end{eqnarray*} 
ordered by logical implication $\models$. The partial order $\langle
\con_\pars,\models\rangle$ is a complete lattice with bottom $\wedge \pars$,
top $1$, meet $\wedge$ and join $\dot{\vee}$ where $ f_1\dot{\vee}
f_2 = \wedge \{f\mid f_1\models f ~\mbox{and}~f_2\models f\}$. The
abstraction and concretization functions are
\begin{eqnarray*} 
\alpha_{\con_\pars}(\Theta) &=& \wedge \{ x \mid x\in \pars ~{and}~\forall \theta\in\Theta.(\vars{\theta(x)}=\emptyset) \} \\
\gamma_{\con_\pars}(\wedge X) &=& \{ \theta \mid \forall x\in{X}.(\vars{\theta(x)}=\emptyset) \} 
\end{eqnarray*} 
$\gc{\wp(\dsub_\pars)}{\alpha_{\con_\pars}}{\con_\pars}{\gamma_{\con_\pars}}$ is a
Galois insertion.

\subsection{Parametrizing Groundness  Analysis} 
  
 A parametric analysis informs about how the abstract property at a
 program point depends on that at an initial program point.  The
 parametric groundness  analysis is obtained by
 parametrizing the abstract interpretation for groundness 
 analysis with the groundness domain $\con_\pars$ where $\pars$ is the set of
 groundness parameters for the variables at the initial program point.
 The primitive abstract domains for the parametric analysis is thus
 $\con_{\pars} \stackrel{m}{\mapsto} \pos_U$ where $U\in\mathcal{U}$. The
 following abstract operations for the parametric analysis are lifted
 from those for the non-parametric groundness  analysis. We
 shall drop the subscript in $\star_{\con_\pars}$. By
 definition~\ref{df:lift},
\begin{eqnarray*} 
\phi_1 (\lift{\wedge}) \phi_2 &=& \lambda g. (\phi_1(g)\wedge \phi_2(g)) \\
\phi_1 (\lift{\vee}) \phi_2 &=& \lambda g. (\phi_1(g)\vee \phi_2(g)) \\
\lift{\pi_X^\sharp} (\phi) &=&  \pi_X^\sharp\circ\phi\\
\lift{\mathcal{R}^\sharp_{\vec{x}\mapsto\vec{y}}}(\phi) &=& \mathcal{R}^\sharp_{\vec{x}\mapsto\vec{y}}\circ\phi
\end{eqnarray*}

\section{Encoding  Parametric Groundness Analysis} \label{sec:encode}
In this section, we encode monotone functions in
$\con_{\pars}\stackrel{m}{\mapsto}\pos_U$ as positive propositional formulas in
$\pos_{U\cup \pars}$. A monotone function $\phi$ is encoded as a formula
$\nabla(\phi)$. This encoding enables us to encode abstract operations on
$\con_{\pars} \stackrel{m}{\mapsto}\pos_U$ in a straightforward manner. It
turns out that the encoding of an abstract operation on $\con_{\pars}
\stackrel{m}{\mapsto}\pos_U$ is exactly the corresponding operation on
$\pos_{U\cup\pars}$.

\paragraph{Encoding of abstract properties}
Let $g\in\con_{\pars}$. Then models of $g$ are closed under
conjunction, that is, $M_1\models g$ and $M_2\models g$ implies
$(M_1\wedge M_2)\models g$ \cite{CortesiFW_JLP96}. Thus, $g$ has a minimum
model which is the conjunction of all its models. The minimum model of 
$g$ is denoted $MM_\pars(g)$. 
\[ MM_\pars(g) = \lambda x\in\pars. \bigwedge \{m(x) \mid m\in (\pars\mapsto\bool)~and~ (m \models g)\} 
\]
Let $BF_{\pars}(m)$ be the propositional formula over propositions in $\pars$ that
has $m$ as its minimal model. The formula is unique modulo logical
equivalence. 
\[ BF_{\pars}(m) = \left(\bigwedge_{u\in \pars, m(u)=1} u\right) \wedge \left(\bigwedge_{u\in \pars, m(u)=0} \neg u\right) 
\]
For instance $BF_{\{u_1,u_2\}}(\{u_1\mapsto 1, u_2\mapsto 0\}) =
u_1\wedge \neg u_2$.

\begin{example} Let $\pars=\{\alpha\}$. 
Then $\con_\pars = \{\alpha,1\}$. $MM_{\pars}(1)=\{\alpha\mapsto 0\}$ and 
$MM_{\pars}(\alpha)=\{\alpha\mapsto 1\}$. Thus,
\( BF_{\pars}(MM_\pars(1)) = \neg \alpha
\) and 
\( BF_{\pars}(MM_\pars(\alpha)) = \alpha
\).
\end{example}

A function $\phi$ from $\con_{\pars}$ to $\pos_{U}$ is
represented as a formula in $\pos_{\pars\cup U}$ via an embedding
function $\nabla$ defined as follows.
\[\nabla(\phi) = \bigvee_{g\in \con_{\pars}} BF_{\pars}(MM_{\pars}(g))\wedge \phi(g) 
\]

\begin{example} Let $\pars=\{\alpha\}$ and $U=\{u\}$ Then $\con_\pars=\{\alpha,1\}$ and $\pos_U=\{u,1\}$. 
There are four  functions from $\con_\pars$ to $\pos_U$:  
\(\phi_1 = \{\alpha\mapsto 1, 1\mapsto 1\}\),
\( \phi_2 = \{\alpha \mapsto u, 1\mapsto 1\}\),
\( \phi_3= \{\alpha\mapsto u, 1\mapsto u\}\) and 
\( \phi_4 = \{\alpha \mapsto 1, 1\mapsto u\}
\).
The first three functions are monotone and the last one is not. The embedding of the three monotone functions are as follows. 
\begin{eqnarray*}
\nabla(\phi_1) &=& 1\\
\nabla(\phi_2) &=& (\alpha \wedge u)\vee ((\neg\alpha) \wedge 1) = (\alpha\rightarrow u)\\
\nabla(\phi_3) &=& u
\end{eqnarray*}
Applying $\nabla$ to $\phi_4$, we obtain \( \nabla(\phi_4) = \alpha \vee u
\).  
The following diagram shows $\pos_{\pars\cup U}$ and encoding of
monotone functions $\con_\pars\mono \pos_U$ via $\nabla$. 
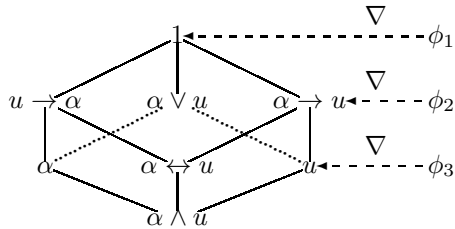
\begin{diagram}[height=1.0em,width=2.5em,abut]
 & & 1 & &  &\lDashto^\nabla &  \phi_1 \\
 & \ruLine & \dLine & \luLine & &&\\
u\rightarrow\alpha  &  & \alpha\vee u  & &\alpha\rightarrow u & \lDashto^\nabla& \phi_2 \\
 \dLine & \ruLine[dotted] \luLine &    & \ruLine \luLine[dotted]  &  \dLine &&\\
\alpha &  & \alpha\leftrightarrow u & & u &\lDashto^\nabla& \phi_3\\
&\luLine&\uLine&\ruLine& && \\
&& \alpha\wedge u & & &  & \\ 
\end{diagram} 
 There are positive propositional formulas in $\pos_{\pars\cup U}$ such as
 $\nabla(\phi_4)$ that are not images of monotone functions in
 $\con_\pars\mono\pos_U$ under $\nabla$. These
 formulas are not used in the parametric analysis.
\end{example}

\begin{lemma} $\nabla$ is  monotone and 1-1. 
{
\begin{proof} 
That $\nabla$ is monotone follows from its definition
straightforwardly.  We now prove that $\nabla$ is 1-1.  Let $\phi_1\neq
\phi_2$. Then there is $g$ such that $\phi_1(g)\neq \phi_2(g)$ implying there
is a truth assignment $m:U\mapsto\bool$ such that $m(\phi_1(g))\neq
m(\phi_2(g))$.   Construct a truth assignment $m':\pars\cup U\mapsto\bool$ as follows. 
\[ m'(y) = \left\{\begin{array}{lr} 
MM_\pars(g)(y) & \mbox{if $y\in \pars$}\\
m(y) & \mbox{otherwise} \end{array} \right.
\]
Then $m'(\nabla(\phi_1)) = m'(\phi_1(g)) = m(\phi_1(g))$ since $\phi_1(g)$ does
not contain any propositional variable in $\pars $. Similarly,
$m'(\nabla(\phi_2)) = m'(\phi_2(g)) = m(\phi_2(g))$. Thus, $\nabla(\phi_1)\neq
\nabla(\phi_2)$. \qed
\end{proof}
}
\end{lemma}

\paragraph{Decoding abstract properties and instantiating analysis}
Since $\nabla$ is 1-1, its inverse exists. Define $\nabla^{-1}(h) =
\lambda g. MM_\pars(g)(h)$. The following lemma proves that $\nabla^{-1}$
is the inverse of $\nabla$.
\begin{lemma} 
$\nabla^{-1}(\nabla(\phi)) = \phi$ for any function in $\con_\pars\mono
  \pos_U$. 
{
\begin{proof} Note that $MM_\pars(g)(BF_\pars(MM_\pars(g'))) = 0$ for any $g'\neq g$. Hence,
$(\nabla^{-1}\circ\nabla(\phi))(g) = MM_\pars(g)(\nabla(\phi)) = \phi(g)$. \qed
\end{proof}
}
\end{lemma}
Instantiating an analysis result $\nabla(\phi)$ for a given input $g$
amounts to calculating $\phi(g)$ which{, according to the above proof,}
amounts to calculating $MM_\pars(g)(\nabla(\phi))$. Thus, instantiating an
analysis result for a given input $g$ does not requires a complete
decoding.

\paragraph{Encoding Analysis Input} 
Let $V$ be the set of variables in the initial goal.  The parametric
analysis can be performed with any monotone function $\con_\pars\mono
\pos_V$ as input. A more natural input associates each variable in $V$
with a different parameter since it allows the influence of the
groundness of the variables in the initial goal on groundness
dependencies at other program points to be observed.  The following
shows that the input has a natural encoding. Define $BM_\pars(X) =
BF_\pars(MM_\pars(\bigwedge X))$ for any $X\subseteq\pars$.

\begin{theorem} 
Let $|V|=|\pars|$, $\rho:\pars \mapsto V$ an invertible function and
$\iota:\con_\pars\stackrel{m}{\mapsto}\pos_V$ defined $\iota(\bigwedge
X) = \bigwedge\{\rho(x)\mid x\in X\}$.  Then $\nabla(\iota) =
\bigwedge_{\alpha\in \pars} (\alpha\rightarrow \rho(\alpha))$.

{
\begin{proof} The proof is by induction on $|\pars|$. 
\begin{itemize} 
\item [Basis]. The thesis holds vacuously for the case $|\pars|=0$. 
\item [Induction]. Assume that thesis holds for all $\pars$ such that $|\pars|=n$. Assume that $|\pars'|=n+1$. There are $\alpha$ and $\pars$ such that $\pars'=\pars\cup\{\alpha\}$ and $\alpha\not\in \pars$. 

\begin{eqnarray*} 
\nabla(\iota) &=& \bigvee_{g\in\con_{\pars'}} BF_{\pars'}(MM_{\pars'}(g))\wedge \iota(g)\\
              &=& \bigvee_{X\in\wp({\pars'})} BM_{\pars'}(X)\wedge \iota(\bigwedge X)\\
              &=& \bigvee_{X\in\wp({\pars})} BM_{\pars'}(X)\wedge \iota(\bigwedge X) \vee \bigvee_{X\in\wp({\pars})} BM_{\pars'}(X\cup\{\alpha\})\wedge \iota(\alpha\wedge\bigwedge X) \\
              &=& (\neg\alpha)\wedge\bigvee_{X\in\wp({\pars})} BM_{\pars}(X)\wedge \iota(\bigwedge X) \vee \alpha\wedge\rho(\alpha)\bigvee_{X\in\wp({\pars})} BM_{\pars}(X)\wedge \iota(\bigwedge X) 
\\
              &=& (\alpha\rightarrow \rho(\alpha))\wedge\bigvee_{X\in\wp({\pars})} BM_{\pars}(X)\wedge \iota(\bigwedge X)\\ 
              &=& (\alpha\rightarrow \rho(\alpha))\wedge\bigwedge_{\beta\in \pars} 
(\beta\rightarrow\rho(\beta)) ~\mbox{by the induction hypothesis}\\
              &=& \bigwedge_{\beta\in \pars'} 
(\beta\rightarrow\rho(\beta))
\end{eqnarray*}
Hence the thesis holds for $\pars'$. \qed
\end{itemize} 
\end{proof}  
}
\end{theorem}  
 
\begin{example} Let $U=\{x_1,x_2\}$ and $\pars=\{\alpha_1,\alpha_2\}$. The encoding of  the monotone function $\{\alpha_1\mapsto x_1, \alpha_2\mapsto x_2, \alpha_1\alpha_2\mapsto x_1\wedge x_2, 1\mapsto 1\}$ is $(\alpha_1\to x_1)\wedge(\alpha_2\to x_2)$. 
\end{example}

\paragraph{Encoding abstract operations}
The encoding $\nabla$ allows us to use the same set of the operations
for both non-parametric and parametric groundness  analyses,
which is formally stated in the following theorem.  The theorem also
states that $\nabla(\con_\pars\stackrel{m}{\mapsto} \pos_X)$ is closed
under all the analysis operations.

\begin{theorem} Let $\pars$ be a set of parameters, $U,V\in\mathcal{U}$, $\vec{x}$ and $\vec{y}$ be vectors of variables such that $|\vec{x}|= |\vec{y}|$. Then 
\begin{enumerate} 
\item $\nabla(\phi_1 (\lift{\wedge}) \phi_2) =
  \nabla(\phi_1)\wedge\nabla(\phi_2)$ for any $\phi_1\in (\con_{\pars}
  \mono \pos_U)$ and $\phi_2\in (\con_{\pars} \mono \pos_V)$;
\item $\nabla( \phi_1(\lift{\vee}) \phi_2 ) = \nabla(\phi_1)\vee\nabla(\phi_2)$ 
for any $\phi_1,\phi_2\in (\con_{\pars} \mono \pos_U)$; 
\item $\nabla(\lift{\pi_X^\sharp}(\phi)) = \pi_X^\sharp(\nabla(\phi))$ for any $\phi\in (\con_{\pars} \mono \pos_U)$ and any $X\subseteq U$; 
\item $\nabla(\lift{\mathcal{R}^\sharp_{\vec{x}\mapsto \vec{y}}}(\phi)) =
  \mathcal{R}^\sharp_{\vec{x}\mapsto\vec{y}}(\nabla(\phi))$ for any $\phi\in
  (\con_{\pars} \mono \pos_U)$.
\end{enumerate} \label{th}

{
\begin{proof} 
 Consider (1) first. Note that $BM_\pars(X)\wedge BM_\pars(Y)=0$ when $X\neq Y$. 
\begin{eqnarray*} 
\nabla(\phi_1)\wedge\nabla(\phi_2) 
&=& (\bigvee_{X\in\wp(\pars)} BM_\pars(X)\wedge \phi_1(\bigwedge X)) \wedge (\bigvee_{Y\in\wp(\pars)} BM_\pars(Y)\wedge \phi_2(\bigwedge Y)) \\
&=& \bigvee_{X\in\wp(\pars), Y\in\wp(\pars)} BM_\pars(X)\wedge BM_\pars(Y)\wedge
 \phi_1(\bigwedge X) \wedge \phi_2(\bigwedge Y) \\
&=& \bigvee_{X\in\wp(\pars)} BM_\pars(X)\wedge \phi_1(\bigwedge X) \wedge \phi_2(\bigwedge X) \\
&=& \nabla(\lambda g. \phi_1(g) \wedge \phi_2(g)) \\
&=& \nabla(\phi_1 (\lift{\wedge}) \phi_2)
\end{eqnarray*}
The proof of (2) is similar. (3) and (4) are straightforward. \qed
\end{proof}
}
\end{theorem}

The following theorem shows that encoding of a monotone function
$\phi$ is logically equivalent to $\bigwedge_{g\in\con_\pars} (g\rightarrow \phi(g))$. 
 
\begin{theorem} 
For any $\pars,V$ such that $\pars\cap V=\emptyset$,
\begin{equation} 
 \forall \phi\in \con_\pars\stackrel{m}{\mapsto} \pos_V. 
\left(\bigvee_{X\in\wp(\pars)} BM_{\pars}(X) \wedge \phi(\bigwedge X) 
   = \bigwedge_{X\in\wp(\pars)} (\bigwedge X \rightarrow \phi(\bigwedge X))
\right) 
\label{eq:equality}
\end{equation}

{
\begin{proof} The proof is done by induction on cardinality of $\pars$.  
\begin{itemize} 
\item [Basis.] $|\pars| = 0$ and hence $\pars=\emptyset$. Then
  $\wp(\pars)=\{\emptyset\}$.  Thus, $\bigvee_{X\in\wp(\pars)} BM_{\pars}(X)
  \wedge \phi(\bigwedge X) = BM_\emptyset(\emptyset) \wedge (\bigwedge
  \emptyset) = \phi(1)$.  We also have $\bigwedge_{X\in\wp(\pars)}
  (\bigwedge X \rightarrow \phi(\bigwedge X)) = \bigwedge\emptyset
  \rightarrow \phi(\bigwedge\emptyset) = \phi(1)$. Hence,
  formula~\ref{eq:equality} holds for the base case.
\item [Induction.] Assume that formula~\ref{eq:equality} holds for any
  $\pars$ such that $|\pars|=n$ and $\pars\cap V=\emptyset$. Let $\pars'=\pars\cup\{z\}$
  where $z\not\in \pars$ is an arbitrary variable and $\phi$ be an arbitrary
  monotone in $\con_{\pars'}\stackrel{m}{\mapsto}\pos_V$. Then
  $|\pars'|=n+1$. Note that $\wp(\pars')=\wp(\pars)\cup \{Y\cup\{z\} \mid
  Y\in\wp(\pars)\}$.
Then
\begin{eqnarray*}
\lefteqn{\bigwedge_{X\in\wp(\pars')} (\bigwedge X \rightarrow \phi(\bigwedge X))}\\
& = & \left(\bigwedge_{X\in\wp(\pars)} (\bigwedge X \rightarrow \phi(\bigwedge X))\right) ~~\wedge~~  \left( \bigwedge_{Y\in\wp(\pars)} (z\wedge\bigwedge Y \rightarrow \phi(z\wedge\bigwedge Y)) \right)\\
& = &  \left(\bigwedge_{X\in\wp(\pars)} (\bigwedge X \rightarrow \phi(\bigwedge X))\right) ~~\wedge~~ \left(\neg z \vee \bigwedge_{Y\in\wp(\pars)} (\bigwedge Y \rightarrow \phi(z\wedge\bigwedge Y)) \right)\\
& = &  \left(\bigwedge_{X\in\wp(\pars)} (\bigwedge X \rightarrow \phi(\bigwedge X))\right) ~~\wedge~~ \left(\neg z \vee \bigwedge_{Y\in\wp(\pars)} (\bigwedge Y \rightarrow \phi'(\bigwedge Y)) \right)
\end{eqnarray*} 
where $\phi'(\bigwedge Y)= \phi(z\wedge \bigwedge Y)$ for all $Y\in\wp(\pars)$. 
Since $\phi\in \con_{\pars'}\stackrel{m}{\mapsto}\pos_V$ and $z\not\in \pars$,
both $\phi \in \con_{\pars}\stackrel{m}{\mapsto}\pos_V$ and $\phi'\in
\con_{\pars}\stackrel{m}{\mapsto}\pos_V$. By applying the induction hypothesis twice, we have
\begin{eqnarray*} 
\lefteqn{\bigwedge_{X\in\wp(\pars')} (\bigwedge X \rightarrow \phi(\bigwedge X))}\\
& = & ( \bigvee_{X\in\wp(\pars)} BM_\pars(X) \wedge \phi(\bigwedge X) ) 
  ~~\wedge~~  (\neg z \vee \bigvee_{Y\in\wp(\pars)} ( BM_\pars(Y) \wedge \phi'(\bigwedge Y)) )\\
& = & \left( \begin{array}{c} 
   \bigvee_{X\in\wp(\pars)} (\neg z\wedge BM_\pars(X) \wedge \phi(\bigwedge X)) \\
  \vee\\
   \bigvee_{X\in\wp(\pars),Y\in\wp(\pars)} ( BM_\pars(X)\wedge BM_\pars(Y) \wedge \phi(\bigwedge X)\wedge \phi'(\bigwedge Y))
\end{array}\right) \\
\end{eqnarray*} 
Since $z\not\in \pars$, $\neg z \wedge BM_{\pars} (X) = BM_{\pars\cup\{z\}}(X) =
BM_{\pars'}(X)$ for any $X\in\wp(\pars)$. Suppose $X,Y\in\wp(\pars)$ and $X\neq
Y$.  Then there is a $v\in \pars$ such that (i) $v\in X\setminus Y$ or
(ii) $v\in Y\setminus X$. Consider the case (i) and let
$X_v=X\setminus\{v\}$ and $\pars_v= \pars\setminus\{v\}$.  $BM_\pars(X)\wedge
BM_{\pars}(Y)= \neg(\bigvee (\pars\setminus X)) \wedge \bigwedge X \wedge \neg
(\bigvee (\pars\setminus Y)) \wedge \bigwedge Y = \neg(\bigvee (\pars\setminus
X)) \wedge \bigwedge X_v \wedge v \wedge \wedge \neg v \neg (\bigvee
(\pars_v\setminus Y)) \wedge \bigwedge Y = 0$.  Similarly,
$BM_\pars(X)\wedge BM_{\pars}(Y)= 0$ in the case (ii). 
By monotonicity of $\phi$, $\phi(\bigwedge X)\wedge \phi'(\bigwedge X) = 
\phi(\bigwedge X)\wedge \phi(z\wedge \bigwedge X) = \phi(z\wedge \bigwedge X)$ 
and $\phi(\bigwedge X)\vee \phi(z\wedge \bigwedge X) = \phi(\bigwedge X)$
for any 
$X\in\wp(\pars)$. Then, 
\begin{eqnarray*} 
\lefteqn{\bigwedge_{X\in\wp(\pars')} (\bigwedge X \rightarrow \phi(\bigwedge X))}\\
& = & 
   \bigvee_{X\in\wp(\pars)} ( BM_{\pars'}(X) \wedge \phi(\bigwedge X)) 
  \vee
   \bigvee_{X\in\wp(\pars)} ( BM_\pars(X) \wedge \phi(\bigwedge X)\wedge \phi'(\bigwedge X))\\
& = & 
   \bigvee_{X\in\wp(\pars)} ( BM_{\pars'}(X) \wedge \phi(\bigwedge X)) 
  \vee
   \bigvee_{X\in\wp(\pars)} ( BM_\pars(X) \wedge \phi(z\wedge\bigwedge X))\\
& = & 
   \bigvee_{X\in\wp(\pars)} ( BM_{\pars'}(X) \wedge \phi(\bigwedge X)) 
  \vee
\bigvee_{X\in\wp(\pars)} (\neg z\wedge BM_\pars(X) \wedge \phi(z\wedge\bigwedge X))\\
&&  
\hspace{10.6pc}  \vee     \bigvee_{X\in\wp(\pars)} (z\wedge BM_\pars(X) \wedge \phi(z\wedge\bigwedge X))\\
& = & 
   \bigvee_{X\in\wp(\pars)} ( BM_{\pars'}(X) \wedge \phi(\bigwedge X)) 
  \vee
 \bigvee_{X\in\wp(\pars)} ( BM_{\pars'}(X) \wedge \phi(z\wedge\bigwedge X))\\
&&  
\hspace{10.6pc}  \vee    \bigvee_{X\in\wp(\pars)} ( BM_{\pars'}(X\cup\{z\}) \wedge \phi(\bigwedge (X\cup\{z\})))\\
& = &    \bigvee_{X\in\wp(\pars)} ( BM_{\pars'}(X) \wedge (\phi(\bigwedge X)\vee \phi(z\wedge\bigwedge X) )) \\
&& \hspace{10.6pc}   \vee
   \bigvee_{X\in\wp(\pars)} ( BM_{\pars'}(X\cup\{z\}) \wedge \phi(\bigwedge (X\cup\{z\})))\\
& = & 
   \bigvee_{X\in\wp(\pars)} ( BM_{\pars'}(X) \wedge \phi(\bigwedge X)) \vee
   \bigvee_{X\in\wp(\pars)} ( BM_{\pars'}(X\cup\{z\}) \wedge \phi(\bigwedge (X\cup\{z\})))\\
&=& \bigvee_{X\in\wp(\pars')} ( BM_{\pars'}(X) \wedge \phi(\bigwedge X)) 
\end{eqnarray*} 
\qed
\end{itemize}
\end{proof}
}
\end{theorem}

\section{Prototype Implementation} \label{sec:imp}
We have implemented a logic program analyzer in SICSTus Prolog and the
CUDD package that can perform both parametric and non-parametric
groundness  analysis. The analyzer takes a text file as
input that contains a Prolog program, a directive of the form
\texttt{:- main(Pred/Arity)} specifying a top-level goal and a
directive \texttt{:- parametric(yes)} if the parametric analysis is to
be performed.

\subsection{Analysis Process} 

The analyzer first does the magic transformation~\cite{DebrayR94} on
the source program and the top-level goal $q(x_1,\cdots, x_n)$ that is
constructed from the directive \texttt{:-main(q/n)}. For each
predicate $p/n$ in the source program, the transformed program
contains two predicates $call\_p/n$ and $ans\_p/n$ such that success
sets for $call\_p/n$ and $ans\_p/n$ are the set of calls to $p$ and
the set of successes of $p$ in the source program during the execution
of the top-level goal. In the second step, the analyzer constructs a
call-graph which captures dependencies between the clauses of the
transformed program and computes strongly connected components (SCCs)
of the call-graph. The third step normalizes the transformed program
and then abstractly compiles~\cite{HermenegildoWD92} the normalized
program by replacing each built-in with its success pattern. For
instance, $x_1=x_3$ is replaced with $x_1\leftrightarrow x_3$. Then,
the unit clause $call\_q(x_1,\cdots,x_n)$ is added for the
non-parametric analysis or the clause $call\_q(x_1,\cdots,x_n)
\mbox{:-} (\beta_1\to x_1) \wedge\cdots\wedge(\beta_n\to x_n)$ is
added otherwise. Then the success pattern of the abstract program is
computed according to the SCCs which yields call and success patterns
for the source program and the top level goal. Note that SCCs are
computed before abstract compilation. This is because abstract
compilation incurs loss of concrete information, which may result in
more dependencies between clauses.

Consider the reverse program with top-level goal $r(x_1,x_2)$. Suppose
that we want to perform the parametric analysis. Then the text file
contains.
\begin{eqnarray}
&\mbox{:-}&main(r/2).\\
&\mbox{:-}&parametric(yes).\\
r([],[]). && \\
r([x_1|x_2],x_3)  &\mbox{:-}& r(x_2,x_4), a(x_4,[x_1],x_3).\\
a([],x,x). && \\
a([x_1|x_2],x_3,[x_1|x_4])  &\mbox{:-}& a(x_2,x_3,x_4).
\end{eqnarray}
The following is the abstract program that is obtained\comment{ from
  magic transformation, normalization and abstract compilation } where
$xy$ abbreviates $x\wedge y$.

\begin{eqnarray}
call\_r(x_1,x_2) &\mbox{:-}& (\beta_1\rightarrow x_1) \wedge 
                             (\beta_2\rightarrow x_2).\\
ans\_r(x_1,x_2)  &\mbox{:-}& call\_r(x_1,x_2) , x_1 x_2.\\
call\_r(x_4,x_5)  &\mbox{:-}& call\_r(x_1,x_2) , (x_1\leftrightarrow x_3 x_4). \\
call\_a(x_5,x_6,x_2) &\mbox{:-}&  call\_r(x_1,x_2) , (x_1\leftrightarrow x_3 x_4), ans\_r(x_4,x_5) , (x_6\leftrightarrow x_3). \\
ans\_r(x_1,x_2) &\mbox{:-}&  call\_r(x_1,x_2) , (x_1\leftrightarrow x_3 x_4), ans\_r(x_4,x_5) , \nonumber \\
&&   (x_6\leftrightarrow x_3) ,  ans\_a(x_5,x_6,x_2). \\
ans\_a(x_1,x_2,x_3) &\mbox{:-}& call\_a(x_1,x_2,x_3) , x_1 \wedge (x_2\leftrightarrow x_3). \\
call\_a(x_5,x_2,x_6) &\mbox{:-}& call\_a(x_1,x_2,x_3) , (x_1\leftrightarrow x_4x_5)\wedge(x_3\leftrightarrow x_4x_6)\\
ans\_a(x_1,x_2,x_3)  &\mbox{:-}&  call\_a(x_1,x_2,x_3) , (x_1\leftrightarrow x_4x_5)\wedge (x_3\leftrightarrow x_4x_6), \nonumber \\
 && ans\_a(x_5,x_2,x_6).
\end{eqnarray}
Each clause in the abstract program is derived from the input file.
The clause 8 results from the clauses 2 and 3, the clause 9 from the
clause 4, the clauses 10,11 and 12 from the clause 5, the clauses 13
from the clause 6 and the clauses 14 and 15 from the clause 7. The
SCCs are $\{8\}, \{9\}, \{10\}$ and $\{11,12,13,14,15\}$ with the
latter SCCs depending only on the earlier ones.  After evaluating the
abstract program, we obtain
\begin{eqnarray*} 
call\_a(x_1,x_2,x_3) & \mbox{:-} & (\beta_1\rightarrow x_1x_2)\\
ans\_a(x_1,x_2,x_3) & \mbox{:-} & (\beta_1\rightarrow x_1x_2) \wedge (x_3\leftrightarrow x_1x_2)\\
call\_r(x_1,x_2) & \mbox{:-} & (\beta_1\rightarrow x_1)\\
ans\_r(x_1,x_2) & \mbox{:-} & (x_1\leftrightarrow x_2) \wedge 
( (\beta_1\vee \beta_2) \to x_1x_2)
\end{eqnarray*}
The call pattern for $r/2$ states that $r/2$ is (recursively) called
with the first argument being a ground term if the first argument of
the top-level goal is ground ($\beta_1=1$).  There is no similar
relationship between the second argument of a recursive call to $r/2$
with the second argument of the top-level goal. This is precise since
$r/2$ is recursively called with its second argument being a fresh
variable in the second clause for $r/2$. The success pattern for
$r(x_1,x_2)$ has two parts. The first part $x_1\leftrightarrow x_2$
is what a goal independent analysis infers and it states that upon
success, $x_1$ is ground iff $x_2$ is. The second part captures the
effect of the groundness parameters on the groundness of the arguments
of the calls. It states that both $x_1$ and $x_2$ are ground if either
argument of the top level goal is ground.

\subsection{An Example} \label{sec:example}
The following is  the quicksort program plus analysis directives. 
The first directive indicates the top-level goal $qs(x_1,x_2)$ and the
second the parametric analysis. Thus, the input abstract property is
$(\beta_1\to x_1)\wedge(\beta_2\to x_2)$.
{\small
\begin{verbatim}
:- main(qs/2).
:- parametric(yes).

app([],L,L).
app([X|L1],L2,[X|L3]) :-  app(L1,L2,L3).

pt([X|T],P,[X|B],A) :- leq(X,P), pt(T,P,B,A).
pt([X|T],P,B,[X|A]) :- gt(X,P),  pt(T,P,B,A).
pt([],_,[],[]).

leq(X,Y) :- X =< Y. 
gt(X,Y) :- X > Y. 

qs([],[]).
qs([X|Xs],Ys) :- pt(Xs,X,U,V), qs(U,S), qs(V,L), app(S,[X|L],Ys).
\end{verbatim}
}
The predicates \texttt{leq/2} and \texttt{gt/2} have been added to
observe the effect of groundness parameters on their arguments. The
following is the analysis result that has been converted manually to
more readable form.
\begin{eqnarray}
call\_gt(x_1,x_2) &\mbox{:-}& \beta_1 \to x_1x_2 \label{eq:callgt}\\
ans\_gt(x_1,x_2) &\mbox{:-}& x_1x_2 \label{eq:ansgt}\\
call\_leq(x_1,x_2) &\mbox{:-}& \beta_1 \to x_1x_2\\
ans\_leq(x_1,x_2) &\mbox{:-}& x_1x_2\\
call\_pt(x_1,x_2,x_3,x_4) &\mbox{:-}& \beta_1\to x_1x_2 \label{eq:callpt}\\
ans\_pt(x_1,x_2,x_3,x_4) &\mbox{:-}& (\beta_1\to x_2)\wedge x_1x_3x_4 \label{eq:anspt}\\
call\_qs(x_1,x_2) &\mbox{:-}& (\beta_1\to x_1)\wedge (\beta_2\to(x_1\vee x_2)) \label{eq:callqs}\\
ans\_qs(x_1,x_2) &\mbox{:-}& (x_1\leftrightarrow x_2) \wedge 
((\beta_1\vee\beta_2)\to x_1x_2) \label{eq:ansqs}\\
call\_app(x_1,x_2,x_3) &\mbox{:-}& x_1\wedge (\beta_1\to x_2)\wedge(\beta_2\to (x_2\vee x_3)) \label{eq:callapp}\\
ans\_app(x_1,x_2,x_3)  &\mbox{:-}& x_1\wedge(x_2\leftrightarrow x_3)\wedge((\beta_1\vee\beta_2)\to x_2x_3) \label{eq:ansapp}
\end{eqnarray} 

The analysis result gives call and success patterns during the
execution of the top-level goal $qs(x_1,x_2)$ using $\beta_1$ for the
groundness of $x_1$ at the beginning of the execution and $\beta_2$
for that of $x_2$.  By assigning $1$ to $\beta_1$ in the righthand
side of Eq.~\ref{eq:callgt}, we obtain $x_1x_2$, implying that $gt/2$
(hence $>/2$) is always called with ground arguments if the first
argument of the top-level goal is ground. Eq~\ref{eq:ansgt} indicates
$gt/2$ (and $>/2$) always instantiates its arguments to ground
terms. Call and success patterns for $leq/2$ are the same as those for
$gt/2$.  This illustrates that the parametric analysis allows us to
infer a sufficient groundness condition on the top-level goal for the
execution of the program to avoid instantiation
errors~\cite{KingLu02}.  Eq.~\ref{eq:callpt} indicates that if the
first argument of the top-level goal is ground ($\beta_1=1$) then
$pt(x_1,x_2,x_3,x_4)$ is always called with both $x_1$ and $x_2$ being
ground.  Eq~\ref{eq:anspt} says that upon success,
$pt(x_1,x_2,x_3,x_4)$ binds $x_1,x_3$ and $x_4$ to ground terms and it
binds $x_2$ to a ground term if $\beta_1=1$. Observe that $x_2$ may be
any term when $x_1,x_3$ and $x_4$ are all empty lists.

The call pattern in Eq.~\ref{eq:callqs} says that $qs(x_1,x_2)$ is
called with $x_1$ ground if $\beta_1=1$ and that either $x_1$ or $x_2$
is ground if $\beta_2=1$. The success pattern for $qs(x_1,x_2)$ in
Eq.~\ref{eq:ansqs} states that $x_1$ is ground iff $x_2$ is ground and
that both $x_1$ and $x_2$ are ground if either $\beta_1$ or $\beta_2$
is ground. \comment{The first part of the success pattern is exactly
  what a goal independent groundness  analysis infers. This
  coincidence happens because a goal independent analysis is performed
  without any information about how the top-level goal is invoked.}
From Eq.~\ref{eq:callapp}, we can infer that when $app(x_1,x_2,x_3)$
is called, $x_1$ is always ground, and $x_2$ is ground if $\beta_1=1$,
and at least one of $x_2$ and $x_3$ is ground if $\beta_2=1$.  From
Eq.~\ref{eq:ansapp}, one can deduce that upon success of
$app(x_1,x_2,x_3)$, $x_1$ is always ground, $x_2$ is ground iff $x_3$
is ground, and both $x_2$ and $x_3$ are ground if either $\beta_1$ or
$\beta_2$ is $1$.

\subsection{Performance}
The analyzer has been tested with a suite of benchmark programs. The
experiments were done on a 2.33GHz Intel (R) Xeon (R) CPU running
Linux 2.6.24 and SICSTUS Prolog 4.0.3. The CUDD package version is
2.4.1.

Table~\ref{fig:poly} shows data from the experiment. All but the last
row corresponds to a benchmark program.  The first column contains the
name of the program and the second specifies the top level goal. In
the third column is the number of atoms in the abstract program.  The
fourth column is the time in millisecond spent on the parametric
analysis using $(\beta_1\to x_1)\wedge\cdots\wedge(\beta_n\to x_n)$ as
the input abstract property. The fifth column contains the time spent
on the non-parametric analysis which is performed without any input
groundness information.  The last column contains the ratio of the
fourth over the fifth. The last row gives the total size, total times
and the average ratio.

\begin{table}[t]
\begin{center}
\begin{tabular}{|l|r|r|r|r|r|}\hline\hline
Program & Top-Level & Size &  Para &  Non-Para & Ratio\\\hline\hline
ann1&go/1&1570&273.68&271.57&1.00\\\hline
asm&asm\_PIL/2&3589&757.89&754.73&1.00\\\hline
boyer&tautology/1&725&63.68&65.78&0.96\\\hline
cs\_r&pgenconfig/1&1101&146.31&140.52&1.04\\\hline
disj\_r&top/1&682&60.52&57.36&1.05\\\hline
dnf&dnf/2&358&29.47&33.15&0.88\\\hline
ga&test\_ga/2&1349&176.31&166.84&1.05\\\hline
gabriel&main/2&377&23.15&23.15&1.00\\\hline
kalah&play/2&855&74.73&76.31&0.97\\\hline
life&life/4&272&15.26&13.68&1.11\\\hline
meta&interpret/1&201&14.73&11.05&1.33\\\hline
nandc&play/1&486&32.10&31.05&1.03\\\hline
nbody&go/2&1431&125.78&120.00&1.04\\\hline
neural&test/2&755&69.47&70.00&0.99\\\hline
peep&comppeepopt/3&1435&180.52&176.84&1.02\\\hline
press&test\_press/2&1303&241.57&232.10&1.04\\\hline
read&read/2&1686&281.05&272.63&1.03\\\hline
reducer&try/2&1063&137.36&123.68&1.11\\\hline
ronp&puzzle/1&340&19.47&18.42&1.05\\\hline
sdda&do\_sdda/4&788&82.10&84.73&0.96\\\hline
semi&go/2&1351&150.00&149.47&1.00\\\hline
simple\_analyzer&main/1&1537&242.63&238.42&1.01\\\hline
tictactoe&play/1&474&34.73&32.10&1.08\\\hline
tsp&tsp/5&391&30.52&25.78&1.18\\\hline
zebra&zebra/7&259&18.42&10.52&1.75\\\hline
Total& &24378&3281.57&3199.99&1.02\\\hline
\end{tabular}
\end{center}
\caption{\label{fig:poly} Performance Comparison between Parametric and Non-Parametric Analyses}
\end{table}
The table indicates that the prototype parametric groundness analyzer
spends an average of $0.135$ seconds to process one thousand atoms in
the abstract program.  This is an acceptable speed for most logic
programs.  The table shows that the time the parametric analysis takes
is from 0.88 to 1.75 times that the non-parametric analysis takes with
an average of 1.02. This indicates that extra cost is negligible for
performing the parametric analysis which yields more general results,
which is quite surprising and promising.

\section{Related Work} \label{sec:related}
The approach proposed in section~\ref{sec:lift} for parametrizing a
base analysis lifts each primitive abstract domain of the base
analysis to its cardinal power with an exponent over which parameters
range. The cardinal power belongs to the standard Cousot and Cousot's
abstract interpretation theory and was proposed
in~\cite{Cousot:Cousot:79} to capture dependencies between abstract
properties of a concrete entity~\cite{GR95b}. Let
$\gc{D}{\alpha_E}{E^\sharp}{\gamma_E}$ and
$\gc{D}{\alpha_B}{B^\sharp}{\gamma_B}$ be a Galois connection. Then
$\gc{D}{\alpha}{E^\sharp\mono B^\sharp}{\gamma}$ is a Galois
connection where $\alpha = \lambda d. \lambda e. (\alpha_B( d \sqcap_D
\gamma_E(e))$ and $\gamma$ is that induced by $\alpha$. The cardinal
power domain in ~\cite{Cousot:Cousot:79} and the relative reduced
power domain in~\cite{GR95} are refinements of the base domain. In
contrast, we use cardinal power to capture dependency of analysis
output on analysis input. 

Parametric analysis abounds in literature. The following are a few
examples.  Chatterjee et.\ al. present a point-to analysis for typed
object oriented languages~\cite{ChatterjeeRL}. This analysis computes
a summary function for each method that expresses the effect of the
method on the points-to solution. The summary function is parametrized
by symbolic unknown initial values and conditions on these values. The
actual-formal bindings are accounted for when points-to information is
propagated into a method from its callers.  Liang and Harrod uses
symbolic names for memory locations whose addresses may be passed into
a procedure~\cite{LiangH01}. These symbolic names are then used in
point-to graphs which expresses parametrized summary information for
a procedure. The summary information can then be instantiated at
specific call sites by binding the symbolic names.  The escape
analysis by Blanchet~\cite{Blanchet03} is a combination of forward and
backward analysis. The backward analysis computes escape information
for method arguments as a function of the escape information for
method result.  These bespoken analyses were not designed by
parametrizing a base analysis. Abstract properties in these analyses
are functions over parameters; thus it is interesting to study whether
and how they can be designed by parametrizing a base analysis.

In~\cite{lu:polygrd} is a groundness analysis of logic programs that
is also parametrized by a number of groundness parameters.  The
analysis is designed from Jones and Sondergaard's analysis by lazily
evaluating operations on groundness parameters. However, it does not
capture groundness dependencies precisely between variables in the
program compared with the parametric groundness analysis presented in
this paper. Moreover, the extra cost of performing that analysis over
the corresponding non-parametric analysis is 78\% which is
significant.

This paper shows by an example that inference of sufficient groundness
condition for error free execution can be done with a traditional top
down forward analysis framework. One benefit that comes with a top
down analysis is that analysis can be made more precise because of
availability of a top level goal. In~\cite{KingLu02}, a backward
analysis is presented to infer sufficient groundness condition for
error free execution. This is no coincidence since information derived
by a forward analysis can be derived by a backward analysis and vice
versus~\cite{Cousot81,KingLu03}.

$\pos$-based goal-independent groundness analysis enjoys the property
of being condensing~\cite{Jacobs:JLP92,Langen91,ms93}. An analysis $F$
that infers output information $F(\phi)$ from input information $\phi$
is condensing if $F(\phi\sqcap\psi)=F(\phi)\sqcap\psi$ for any $\phi$
and $\psi$. Thus, a condensing analysis can be performed with partial
input information $\phi$ and its output be conjoined with additional
input information $\psi$ to obtain the output that would result from
analyzing the program with complete input information
$\phi\sqcap\psi$.  Condensing has been studied exclusively for goal
independent analysis. Condensing can be used to retrieve abstract
answers but does not precisely keep track of dependencies between a
top level call and a descendant call because the projection operator
discards useful information that is essential for maintaining such
dependencies. 
\begin{example}
Consider the quicksort program in Section~\ref{sec:example}.  A
non-parametric $\pos$-based goal dependent analysis infers
$call\_app(y_1,y_2,y_3)\mbox{:-}y_1$ from analysis input
$call\_qs(x_1,x_2)\mbox{:-}true$ and it infers
$call\_app(y_1,y_2,y_3)\mbox{:-}y_1\wedge y_2$ from analysis input
$call\_qs(x_1,x_2)\mbox{:-} x_1$. The second call pattern $y_1\wedge
y_2$ for \texttt{app}/3 cannot be obtained as the conjunction of the
call pattern $x_1$ for $qs(x_1,x_2)$ in the second analysis input and
the first call pattern $y_1$ for \texttt{app}/3.
\end{example}

\comment
{Parametric analysis is closely related to analysis which use
relational abstract domains.  For instance, Cousot and Halbwachs
\cite{CousotH78} use polyhedra to approximate functions.


}

\section{Conclusion} \label{sec:conc}
We have proposed an approach to parametrizing a base analysis by
lifting its primitive abstract domains to their cardinal powers and
obtained a parametric groundness analysis for logic programs using
this approach. We have also used positive propositional formulas to
encode abstract properties and presented experimental results on a
suite of benchmark programs.  The experiments show that the parametric
groundness analysis is as fast as the non-parametric groundness
analysis from which it is obtained.


\end{document}